\documentstyle[prl,aps,twocolumn,epsfig,floats,times]{revtex}

\draft

\begin{document} 
\twocolumn[\hsize\textwidth\columnwidth\hsize\csname @twocolumnfalse\endcsname

\title{ New High Temperature Diboride Superconductors: AgB$_2$ and AuB$_2$ } 
\author{ S. K. Kwon$^1$, S. J. Youn$^2$, K. S. Kim$^2$, and B. I. Min$^1$ } 
\address{
$^1$ Department of Physics, Pohang University of Science and Technology, 
     Pohang 790-784, Korea \\
$^2$ 
     National Creative Research Initiative Center 
     for Superfunctional Materials, Department of Chemistry, \\
     Pohang University of Science and Technology, 
     Pohang 790-784, Korea}
\date{\today}
\maketitle

\begin{abstract}
Based on electronic structure calculations,
we have found that noble metal diborides, 
AgB$_2$ and AuB$_2$, are potential candidates 
for conventional BCS-type high temperature superconductors. 
B $2p$ density of states dominates at the Fermi level
in comparison with Ag $4d$ and Au $5d$ states.
Furthermore, the electron-phonon coupling constant $\lambda$ is 
much larger in AgB$_2$ and AuB$_2$ than in MgB$_2$. 
Estimated transition temperatures for AgB$_2$ and AuB$_2$ are 
$T_{\rm c}^{\rm cal} =$ 59 K and 72 K, respectively. 
These are about 2 $\sim$ 3 times higher 
than the estimated $T_{\rm c}^{\rm cal} =$ 27 K in MgB$_2$
and almost comparable to those in cuprate superconductors.
\end{abstract}

]
Due to the highest transition temperature of $T_{\rm c} \simeq$ 39 K\cite{Nagam}
among nonoxide bulk metallic superconductors, 
MgB$_2$ has renewed scope of the transition temperature 
in the conventional BCS superconductors. 
Photoemission spectroscopy\cite{Takah}, tunneling spectroscopy\cite{Bolli}, 
isotope effect measurements\cite{Bud'k,Hinks}, and 
inelastic neutron scattering measurements\cite{Osbor} support 
that MgB$_2$ belongs to the $s$-wave phonon-mediated BCS superconductor.
Borons in MgB$_2$ can generate high frequency in-plane phonon modes 
caused by their light atomic masses. 
These phonons are thought to be responsible 
for the observed unusual high $T_{\rm c}$.
It is also found that electron doped Mg$_{1-x}$Al$_x$B$_2$ 
loses superconductivity for $x >$ 0.1\cite{Slusk}.
In MgB$_2$, the Fermi level ($E_{\rm F}$) is located at the shoulder 
of the density of states (DOS) curve in which B $2p$ states, 
believed to become superconducting carriers, are dominant.
Thus, electron doping effect in Mg$_{1-x}$Al$_x$B$_2$ 
would move $E_{\rm F}$ to the lower DOS side
and destroy B $2p$ hole states and superconductivity 
by reducing the effective electron-phonon interaction strength \cite{An,Kortu}.

MgB$_2$ research society becomes rapidly growing
and it is amazing that some significant progresses 
have already been made  
toward technological applications\cite{Kang,Canfi,Eom,Bugos,Jin}.
However, it is open and natural to ask 
how many diborides can be categorized as superconductors 
with $T_{\rm c}$ as high as MgB$_2$  
and what is the upper limit of the transition temperature
in these BCS superconductors.

In hexagonal AlB$_2$-type structure, MgB$_2$ is special 
with superconductivity itself even excluding high $T_{\rm c} \simeq$ 39 K.
Other than MgB$_2$, only a few diborides show superconductivity: 
Zr$_{0.13}$Mo$_{0.87}$B$_2$, NbB$_2$\cite{Coope},
and TaB$_2$\cite{Kaczo} with $T_{\rm c} \lesssim$ 10 K.
There are, however, still controversies 
for NbB$_2$ and TaB$_2$\cite{Gasp1,Singh,Rosne}.
Since the discovery of MgB$_2$, there have been several theoretical 
studies to search for the potential high $T_{\rm c}$ binary and ternary borides 
in isoelectronic systems such as BeB$_2$ and CaB$_2$, transition metal (TM)
diborides $TM$B$_2$, and hole doped systems such as Mg$_{1-x}M_x$B$_2$
($M=$ Li, Na, Cu, and Zn) \cite{Satta,Medve,Ravin,Medve2}. 
However, most of the recent experimental works for related binary 
and ternary borides fail to find another high temperature 
diboride superconductor (HT$_{\rm c}$BS)\cite{Rosne,Zhao,Morit}.
Nevertheless, in the present work, we try to give positive answers to
the given questions.  
We suggest that noble metal diborides, AgB$_2$ and AuB$_2$
which correspond to effectively hole doped systems, are potential
candidates for HT$_{\rm c}$BS. 
We have found that their transition temperatures 
are 2 $\sim$ 3 times higher than that in MgB$_2$. 

\begin{figure}[t]
\epsfig{file=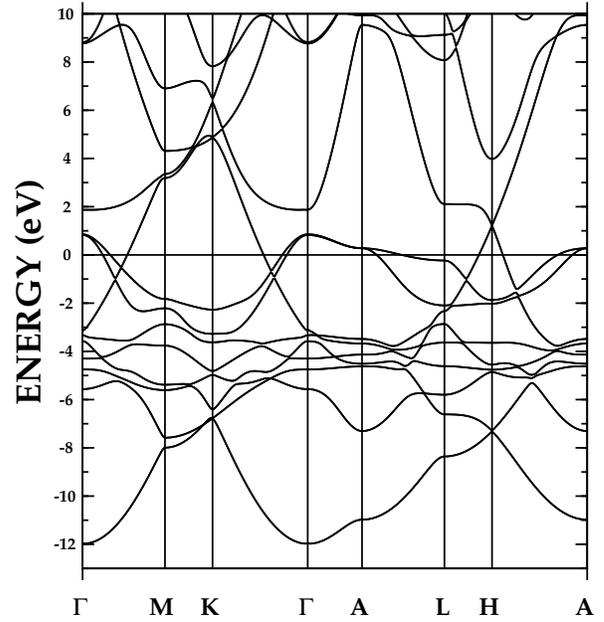,width=8.0cm}
\caption{\label{Agband}
AgB$_2$ band structures along high symmetry lines.
Along $\Gamma-$A$-$L lines, B $2p_{x,y}\sigma$ hole bands are 
formed near $E_{\rm F}$ similarly to MgB$_2$.}
\end{figure}
\begin{figure}[t]
\epsfig{file=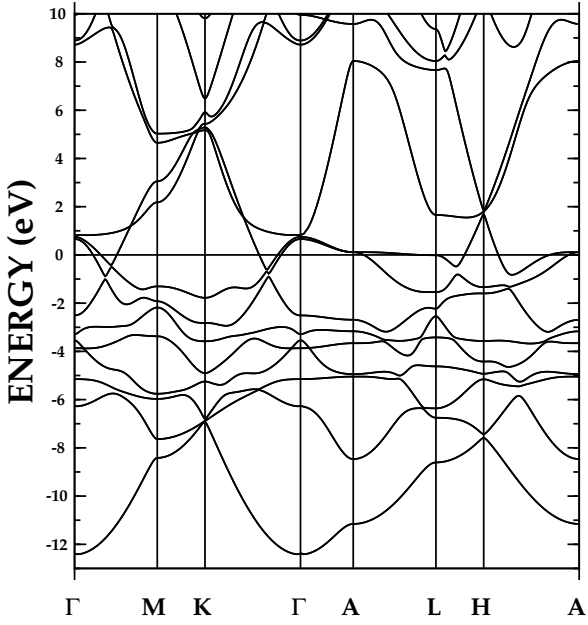,width=8.0cm}
\caption{\label{Auband}
AuB$_2$ band structures along high symmetry lines.
Overall features are similar to those of AgB$_2$}
\end{figure}
\begin{figure}[t]
\epsfig{file=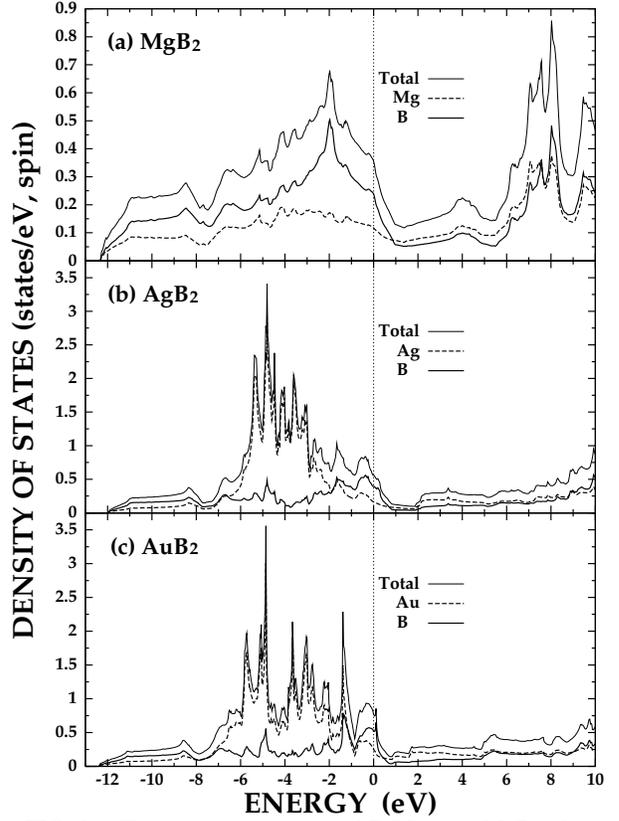,width=8.0cm}
\caption{\label{DOS}
The atomic site projected DOS in (a) MgB$_2$, (b) AgB$_2$, and (c) AuB$_2$.
For the B DOS, two B atomic contributions are summed.
In all cases, the B $2p$ DOS is dominant at $E_{\rm F}$.}
\end{figure}
\begin{figure}[t]
\epsfig{file=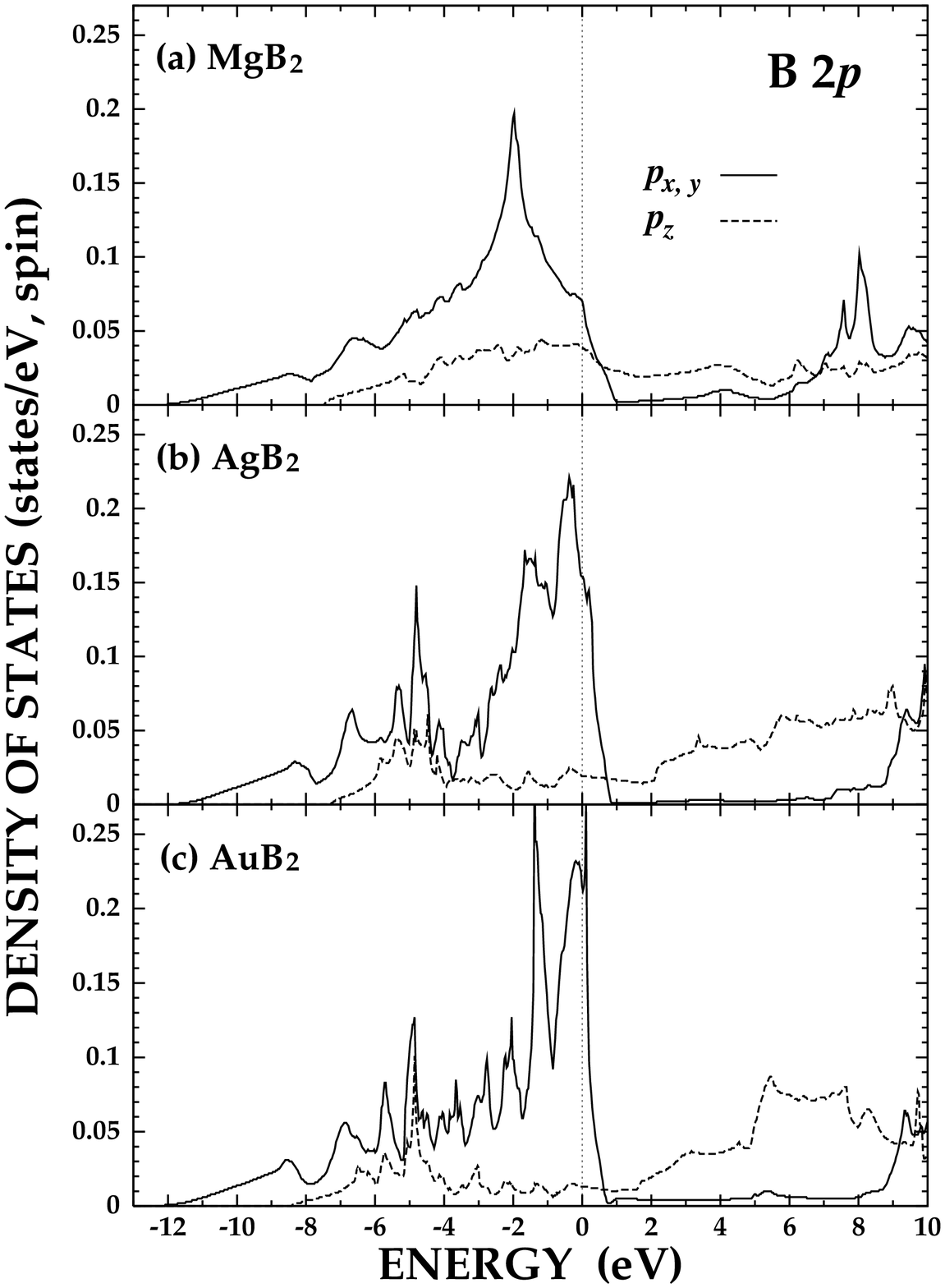,width=8.0cm}
\caption{\label{B2pDOS}
Directionally decomposed B $2p$ DOS in (a) MgB$_2$, (b) AgB$_2$, 
and (c) AuB$_2$.  
In going from MgB$_2$ to AgB$_2$ and AuB$_2$,
the B $2p_{x,y}\sigma$ DOS at $E_{\rm F}$ becomes larger.  }
\end{figure}
To verify its possibility, we have performed fully relativistic 
electronic structure calculations for AgB$_2$ and AuB$_2$ 
by using the linearized muffin-tin orbital (LMTO) band method. 
We have used lattice constants of $a =$ 2.98 \AA~ and $c =$ 3.92 (4.05) \AA~ 
for AgB$_2$ (AuB$_2$)\cite{ysj}.
Hence the B-B interlayer distances are much increased as compared to
that in MgB$_2$, while the B-B in-plane bond lengths are similar.
The Coulomb exchange correlation potential is treated 
in the von Barth-Hedin form of the local spin-density approximation.
For the self-consistent charge density integration, 
280 $k$-points sampling is done in the irreducible wedge 
of the first Brillion zone (BZ).  Basis functions are adopted 
up to $l =$ 3 for Ag and Au sites and $l =$ 2 for B. 

In Fig. \ref{Agband} and Fig. \ref{Auband}, 
we have shown band structures of AgB$_2$ and AuB$_2$,
respectively, along high symmetry lines in the BZ.  
It is clear that B $2p$ states near $E_{\rm F}$,
corresponding to the in-plane B-B $p_{x,y}\sigma$ bands, 
are almost dispersionless along $\Gamma-$A$-$L line
and yield the hole Fermi surfaces.
The effect of the spin-orbit interaction on these bands
is minor: the splittings are barely seen in the case of AuB$_2$.
Previously, similar band feature producing the cylindrical hole Fermi surface
along $\Gamma-$A is emphasized in MgB$_2$, 
as is responsible for superconductivity\cite{An,Kortu}.
However, the B $p\sigma$ bands are flatter in AgB$_2$ and AuB$_2$ 
than in MgB$_2$, yielding the higher DOS at $E_{\rm F}$ 
(Fig. \ref{DOS}).
The bandwidths of B $p\sigma$ are less than $5$ eV in AgB$_2$ and AuB$_2$ 
due to band repulsion between Ag $4d$ (Au $5d$) and B $2p$ 
as is evident along A$-$L line. 
Whereas they are as much as $9$ eV in MgB$_2$ (Fig. \ref{B2pDOS}).  
The band repulsion drives the B $2p\sigma$ sates to pile up at $E_{\rm F}$. 

The calculated DOS's at $E_{\rm F}$, $N(0)$, 
in various systems are summarized in Table \ref{table1}.
We have found that they are about 1.6 and 2.2 times larger
in AgB$_2$ and AuB$_2$ with $N(0) =$ 0.557 and  0.768 states/(eV, spin) 
than $N(0) =$ 0.356 states/(eV, spin) in MgB$_2$.
In addition, the B $2p$ contributions to the total DOS at $E_{\rm F}$ 
are $\tilde N_{\rm B}(0) \simeq$ 0.6 in all cases.
The total and atomic site projected DOS's in Fig. \ref{DOS} 
also reflects these features, 
in which B atoms dominantly contribute to the DOS at $E_{\rm F}$. 

Employing the McMillan's empirical formula for $T_{\rm c}$ \cite{McMil},
\begin{eqnarray}
T_{\rm c} = \frac{\Theta_D}{1.45}
\exp\left[\frac{-1.04(1+\lambda)}{\lambda-(1+0.62\lambda)\mu^{\ast}}\right],
\end{eqnarray}
we have estimated the transition temperatures 
with the Coulomb pseudopotential parameter of $\mu^{\ast} =$ 0.1 
and the the experimental Debye temperature of $\Theta_{D} \sim$  700 K
\cite{Bud'k,Osbor}.
We consider the superconducting properties
within the framework of the simple rigid-ion approximation\cite{Gaspa}.
Calculation of $\eta_{\alpha}=N(0)\langle I^2_{\alpha}\rangle$, 
where $\langle I^2_{\alpha}\rangle$ is the average electron-ion 
interaction matrix element for the $\alpha$ atom in the unit cell, 
yields $\eta_{\alpha} =$  0.03 and 1.87 eV/\AA$^2$ 
for $\alpha =$ Mg and B atoms, respectively.
Note that the contribution from B atom is dominating.
Then the electron-phonon coupling constant, 
$\lambda = \sum_{\alpha}\eta_{\alpha} /M_{\alpha}\langle\omega^{2}\rangle$, 
is evaluated to be $\lambda =$ 0.79. 
Here $M_{\alpha}$ is an ionic mass 
and $\langle\omega^{2}\rangle$ is the average phonon frequency 
approximated by $\langle\omega^{2}\rangle\simeq\Theta_{D}^{2}/2$. 

First of all, we would like to point out that our estimated values 
of $\lambda =$ 0.79 and the corresponding $T_{\rm c}^{\rm cal} =$ 27 K 
for MgB$_2$ are in good agreement with other theoretical\cite{An,Kortu,Kong} 
and experimental values.
In a similar way, we have also estimated superconducting properties
of AgB$_2$ and AuB$_2$, assuming that the phonon frequencies are
of the same order as that in MgB$_2$. 
It is a plausible approximation 
because the ionic radii of Ag and Au are similar to that of Mg\cite{Kitte} 
and the relevant phonon mode is the in-plane B bond stretching mode.
Remarkably, as shown in Table \ref{table1},
we have obtained almost twice larger $\eta_{\rm B}$'s for AgB$_2$ and AuB$_2$
than that for MgB$_2$, and so the electron-phonon coupling constant 
values are $\lambda =$ 1.35 and 1.65 for AgB$_2$ and AuB$_2$, respectively.
With these values, we have estimated the transition temperatures of 
$T_{\rm c}^{\rm cal} =$ 59 K for AgB$_2$ and 72 K for AuB$_2$.
They are 2 $\sim$ 3 times larger than that in MgB$_2$,
and even comparable to those in high temperature cuprate superconductors. 

\begin{table}[t]
\caption{\label{table1}
The DOS's at $E_{\rm F}$, $N(0)$ and $N_{\rm B}(0)$, are 
in unit of states/($M$B$_2$, eV, spin) 
and states/(B $2p$, eV, spin), respectively.
$\tilde N_{\rm B}(0)$ given by the ratio of $2N_{\rm B}(0)/N(0)$ is 
B $2p$ contribution to the total DOS at $E_{\rm F}$. 
$n_{\rm h}$ denotes the number of holes in B-B $p\sigma$ band per B atom.
$\eta_{\rm B}$ is the McMillan-Hopfield parameter for B atom 
and $\lambda$ is the electron-phonon coupling constant. 
The transition temperature $T_{\rm c}^{\rm cal}$ is estimated 
by using the McMillan's formula with $\mu^{\ast}=$ 0.1 and $\Theta_{D}=$ 700 K.
For MgB$_2$, present results are consistent with other calculations.}
\begin{tabular}{lccccccc}
            & $N(0)$&$N_{\rm B}(0)$&$\tilde N_{\rm B}(0)$ &$n_{\rm h}$
            &$\eta_{\rm B}$&$\lambda$  &  $T_{\rm c}^{\rm cal}$ (K)\\ \hline
MgB$_2$       & 0.356 & 0.109        & 0.612 & 0.062 & 1.87 &0.79    & 27   \\
CuB$_2$$^{a}$ & 0.710 & 0.195        & 0.550 & 0.088 & 3.50 &1.48    & 65   \\ 
AgB$_2$       & 0.557 & 0.173        & 0.628 & 0.128 & 3.17 &1.35    & 59   \\
AuB$_2$       & 0.768 & 0.225        & 0.586 & 0.120 & 3.88 &1.65    & 72   \\
\end{tabular}
$^a$ Reference\cite{CuT_c}
\end{table}
It may be difficult to find right condition of synthesizing 
stoichiometric AgB$_2$ and AuB$_2$ compounds.
In such a case, first attempt would be to synthesize hole doped materials 
like Mg$_{1-x}$Ag$_x$B$_2$ and Mg$_{1-x}$Au$_x$B$_2$ or to grow them in films.
At least, hole doped samples could be obtained, considering that 
Mg$_{1-x}$Li$_x$B$_2$ has been already tested including $x = 1.0$\cite{Zhao}.
On the other hand, it was reported that Cu substitution in MgB$_2$ produces 
MgCu$_2$ phase rather than CuB$_2$ \cite{Morit}.
This unfortunate occasion is likely to arise 
because the ionic radius difference between Cu and Mg is large, 
and hence Cu prefers to sit at the B site \cite{CuT_c}.  
On the other hand, Ag and Au share similar ionic radius with Mg, 
and so it is expected that these elements will be in  better situation 
than the Li and Cu cases. 
In addition, one may use Mg$_{1-x}$Ag$_x$B$_2$ and Mg$_{1-x}$Au$_x$B$_2$ 
to examine the existing theory which sets the upper limit of 
the transition temperature in the phonon-mediated BCS mechanism 
superconductor as $T_{\rm c} \lesssim$ 40 K. 
If the proposed noble metal HT$_{\rm c}$BS's are realized, 
it is expected that, besides high $T_{\rm c}$,
they will also have the enhanced hole concentration 
of $n_{\rm h} =$ 0.120 and 0.128 holes/B 
in comparison with $n_{\rm h} =$ 0.06 holes/B in MgB$_2$ 
(Table \ref{table1} and Fig. \ref{B2pDOS}). 
These things will serve more favorably for technological applications. 

In conclusion, we have performed band structure calculations 
for AgB$_2$ and AuB$_2$ to search for new HT$_{\rm c}$BS. 
These are predicted to be very promising materials to meet the purpose,
and thus the material synthesis and experimental measurements are encouraged. 

Acknowledgements$-$
This work was supported by KOSEF through eSSC at POSTECH   
and in part by the BK21 Project.


\begin{thebibliography}{99}
\bibitem{CuT_c} Although CuB$_2$ phase was not produced experimentally, 
                electronic structure calculation is performed 
                for $a =$ 2.96 \AA~ and $c =$ 3.25 \AA~, 
                which are optimized values by using the FLAPW method. 
                The transition temperature 
                ignoring the Cu $d$-$f$ channel in the rigid-ion approximation
                is estimated to be $T_{\rm c} =$ 65 K.
\end{thebibliography}
\end{document}